\documentclass[3p,times,procedia]{elsarticle}
\usepackage{nupha_ecrc}
\usepackage{lineno}

\volume{00}

\firstpage{1}

\journalname{Nuclear Physics A}

\runauth{Nirbhay Kumar Behera}

\jid{nupha}

\jnltitlelogo{Nuclear Physics A}

\usepackage{amssymb}

\usepackage[figuresright]{rotating}

\begin{document}

\begin{frontmatter}



\dochead{XXVIIth International Conference on Ultrarelativistic Nucleus-Nucleus Collisions\\ (Quark Matter 2018)}

\title{Higher moment fluctuations of identified particle distributions
  from ALICE}


\author{Nirbhay Kumar Behera for the ALICE collaboration}

\address{Inha University, 100, Inharo, Namgu, Incheon, South Korea}

\begin{abstract}
Cumulants of conserved charges fluctuations are regarded as a potential tool to study the criticality in the QCD phase diagram and to determine the freeze-out parameters in a model-independent way. At LHC energies, the measurements of the ratio of the net-baryon (net-proton) cumulants can be used to test the lattice QCD predictions. In this work, we present the first measurements of cumulants of the net-proton number distributions up to $4^{th}$ order in Pb--Pb collisions at $\sqrt{s_{\mathrm{NN}}}$ = 2.76 and 5.02 TeV as a function of collision centrality. We compare our cumulant ratios results with the STAR experiment net-proton results measured in the first phase of the Beam Energy Scan program at RHIC. The results can be used to obtain the chemical freeze-out parameters at LHC.
\end{abstract}


\end{frontmatter}



\section{Introduction}
The study of fluctuations of conserved charges in heavy-ion collision
experiments is regarded as an excellent observable to map the QCD
phase diagram. Heavy-ion collision experiments at LHC energies aim to
explore the QCD phase diagram at the chiral limit, where the baryo-chemical potential ($\mu_B$) is very small. According to lattice QCD, the phase transition is a
crossover at vanishing $\mu_B$ \cite{Aoki:2006we}. It has been also proposed that at the chiral
limit of light quarks, the phase transition is of the second order, belonging
to the universality class of 3-dimensional O(4) spin model
\cite{Friman:2011pf}. Based on this model, the chiral crossover transition line
will appear close to the freeze-out line. Therefore, the determination of freeze-out parameters is very
important to locate the phase boundary at LHC energies. The freeze-out parameters are obtained by fitting the
particle yields and then comparing them with thermal statistical
models like the Hadron Resonance Gas (HRG) model \cite{Andronic:2009qf}. Recently, the chemical freeze-out temperature is estimated as 156.5 $\pm$ 1.5 MeV at LHC energy by a statistical hadronization model using the ALICE particle yields \cite{Andronic:2016nof}. On the other hand, the crossover temperature,
$T_{c}$, estimated by lattice QCD calculations is 155 (1) (8) MeV
\cite{Bhattacharya:2014ara}. It can be observed from the thermal model
and from lattice QCD calculations that the freeze-out temperature is close to the crossover temperature. Hence, additional measurements of the freeze-out temperature are needed to test the lattice QCD predictions

Experimentally measured cumulants (C$_{n}$) of conserved charge distributions,
like net-charge and net-baryons can be directly connected with various
quark number susceptibilities ($\chi^{(n)}_{q}$) as $\textnormal{C}_{n} = VT^{3} \chi^{(n)}_{q}$
\cite{Karsch:2010ck}. Here the quark number susceptibilities are given by -

\begin{equation}
\chi^{(n)}_{q} = \frac{\partial^{n}[\mathrm{P}(T,\mu)/T^{4}]}{\partial(\mu_{q}/T)^{n}}~.
\label{cn}
\end{equation}

In Eq. \ref{cn}, P is the pressure, $T$ is the temperature and $V$ is the
volume of the system. To get rid of the volume term in Eq. \ref{cn}, the freeze-out parameters can be determined by taking the ratio of the various orders of cumulants \cite{Friman:2011pf,Karsch:2010ck}.

In this work the cumulants of net-proton number fluctuations up to $4^{th}$ order
in minimum-bias Pb--Pb collisions at $\sqrt{s_{\mathrm{NN}}}$ = 2.76 and 5.02 TeV are reported. The ratio of cumulants, C$_3/$C$_{2}$ and
C$_4/$C$_{2}$ are compared with Skellam expectations as a function of
centrality. They are also compared with RHIC Beam Energy Scan
(BES) results.

\section{Experimental setup and analysis methodology}
The analysis is carried out using data from Pb--Pb collisions recorded
at $\sqrt{s_{\mathrm{NN}}}$ = 2.76 and 5.02 TeV with the ALICE detector
\cite{Aamodt:2008zz}. Around 14$\times 10^{6}$ and 59 $\times 10^{6}$
minimum-bias events are used for this analysis at $\sqrt{s_{\mathrm{NN}}}$ = 2.76 and 5.02 TeV, respectively. The V0 detectors covering the
pseudorapidity ranges 2.8 $< \eta < $ 5.1 (V0A) and -3.7 $< \eta < $
-1.7 (V0C) are used for the trigger and centrality estimation. A
minimum-bias trigger is defined by the coincidence of a signal in both the V0A and V0C. The events are classified into different centrality
classes using the V0 signal amplitudes \cite{Aamodt:2010cz}. To reject
the background events and secondary interactions, only events with a reconstructed primary vertex ($V_{z}$) $<$ 10 cm from the center of the ALICE coordinate system (along the z-direction) are considered. The contribution from pileup events for Pb--Pb collisions $\sqrt{s_{\mathrm{NN}}}$ = 2.76 TeV is negligible, while for 5.02 TeV data the pileup events are removed using the correlation between number of hits in Silicon Pixel Detector (SPD) and the multiplicity of V0 detector. For this analysis, tracks reconstructed using the Time Projection Chamber (TPC) with transverse momentum ($p_{\mathrm{T}}$) range 0.4 $<
p_{\mathrm{T}} < $ 1.0 GeV/$c$ and pseudorapidity range of -0.8 $< \eta <$
0.8 are selected.  To ensure good quality tracks, the following criteria are applied. The reconstructed
tracks are selected with at least 80 out of 159 space points in the TPC. Reconstructed tracks have been selected requiring a good quality fit ($\chi^2/NDF < $ 4). Tracks associated to weak leptonic decay topologies are rejected. Furthermore, a $p_{\mathrm{T}}$
dependent selection criteria based on the distance of closest approach (DCA) is imposed to minimize the contribution from secondaries and weak decays. For the (anti-)proton identification the
specific ionization energy-loss ($dE/dx$) information of a given track
in TPC volume is used. A condition of $|n\sigma |< 2.5$ around the
expected mean values of $dE/dx$ for (anti-)proton is applied. The contamination due to particle misidentification for $p_{\mathrm{T}} < 0.85$ GeV/$c$ is negligible, and for 0.85 $< p_{\mathrm{T}} < $ 1.0 GeV/$c$ is around
10$\%$. This effect is taken into account in the systematic uncertainties.

The $p_{T}$ dependent proton (p) and anti-proton ($\bar{\mathrm{p}}$) reconstruction efficiencies are estimated by means of a full Monte Carlo simulation, using the HIJING event generator and the GEANT3 transport code, followed by the standard reconstruction procedure \cite{Wang:1991hta,Brun:1994aa}. The reconstruction efficiency of p and $\bar{\mathrm{p}}$ are about 65$\%$ and $60\%$, respectively. The efficiency correction of the $\langle \mathrm{p}
\rangle$, $\langle \bar{\mathrm{p}} \rangle$ numbers and the cumulants of net-proton distributions
are done using the method proposed in Ref. \cite{Nonaka:2017kko}. The statistical uncertainties are estimated using
the Subsample method with 30 subsamples. The systematic uncertainties are estimated by varying the track quality and
$V_z$ selection criteria with respect to the default ones. The final results reported here are corrected for the centrality bin width  \cite{Luo:2013bmi}.

\section{Results and discussion}

\begin{figure}[h]
\begin{center}$
\begin{array}{ccc}
\includegraphics[width=1.8in]{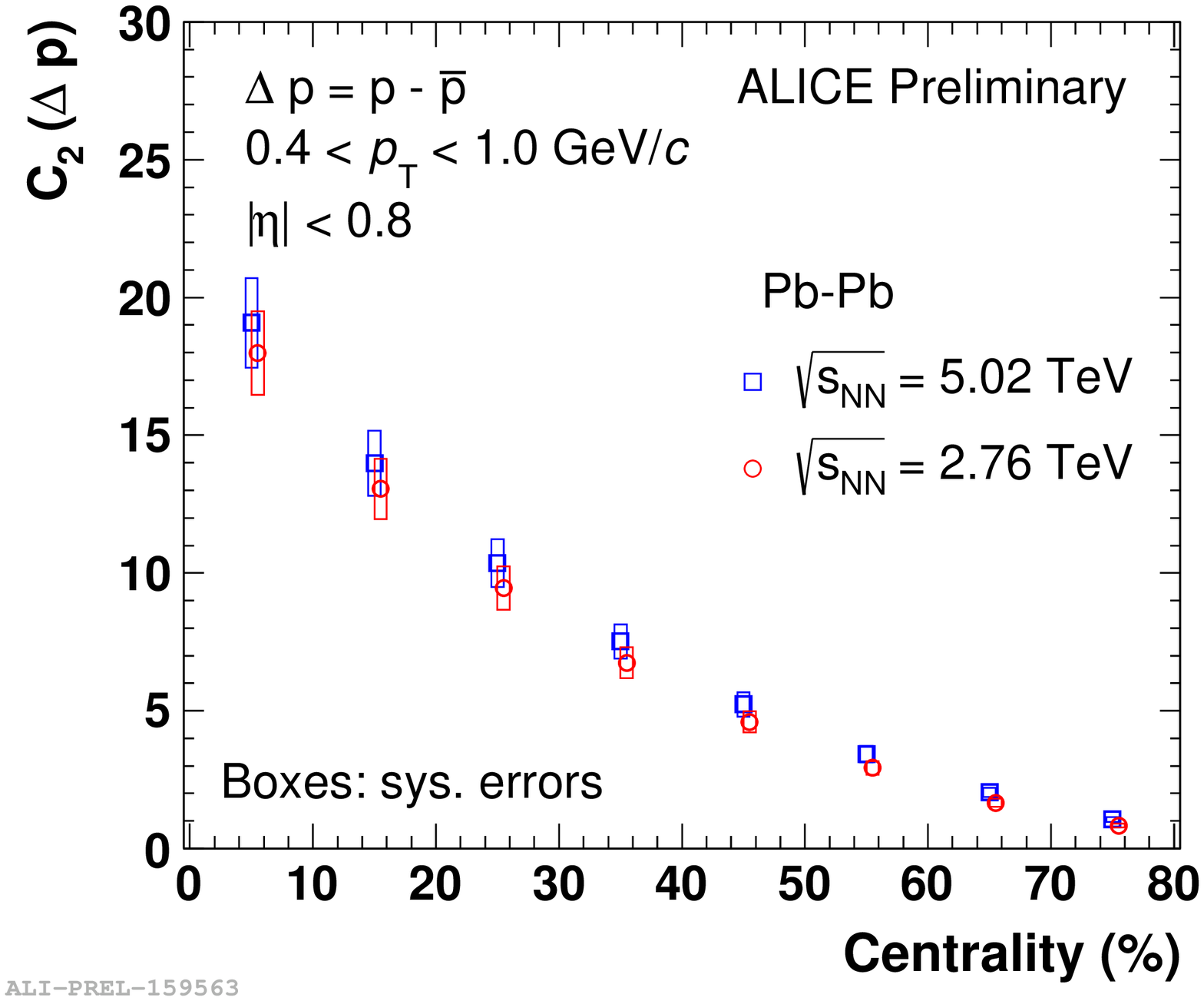} &
\includegraphics[width=1.8in]{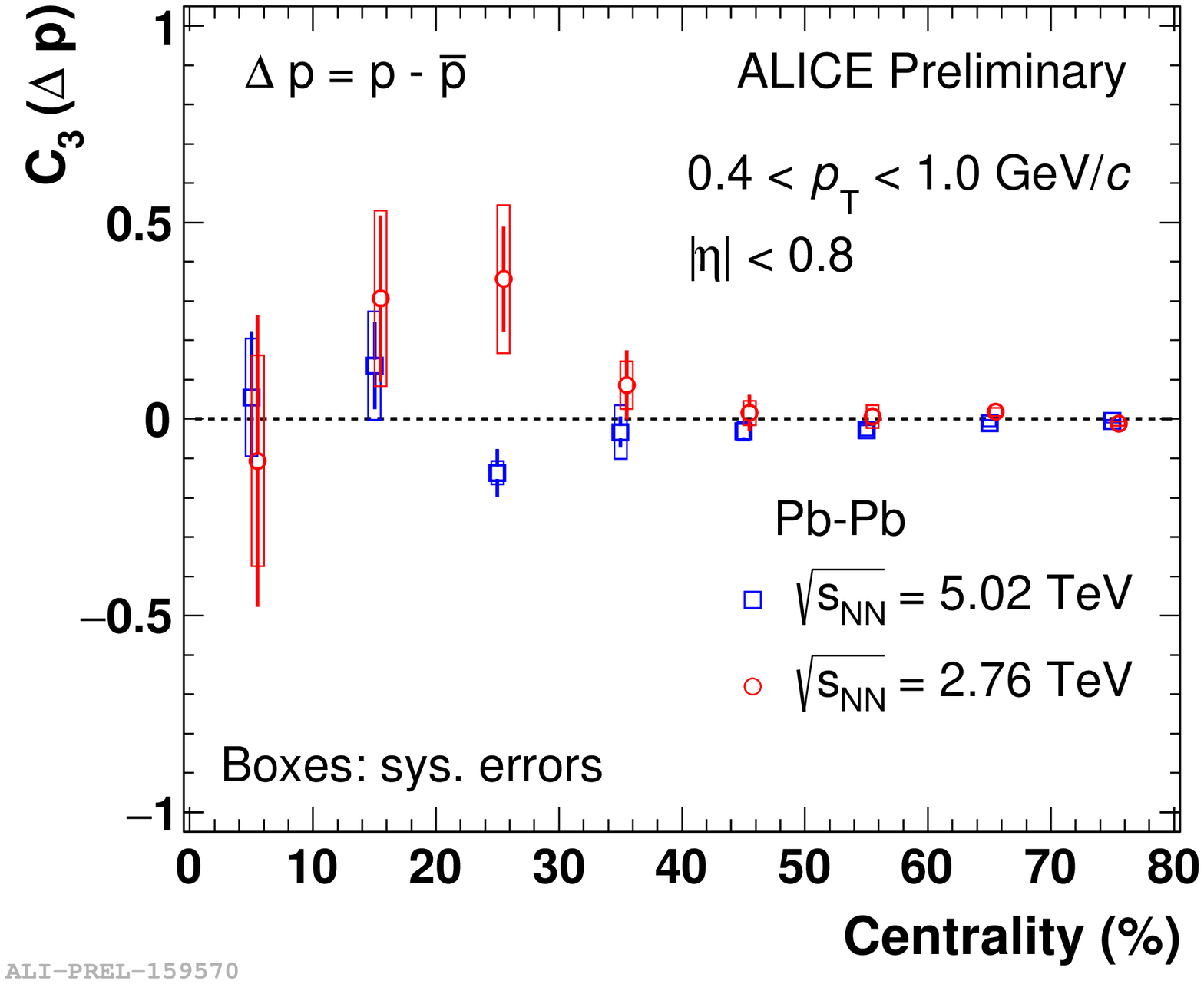} &
\includegraphics[width=1.8in]{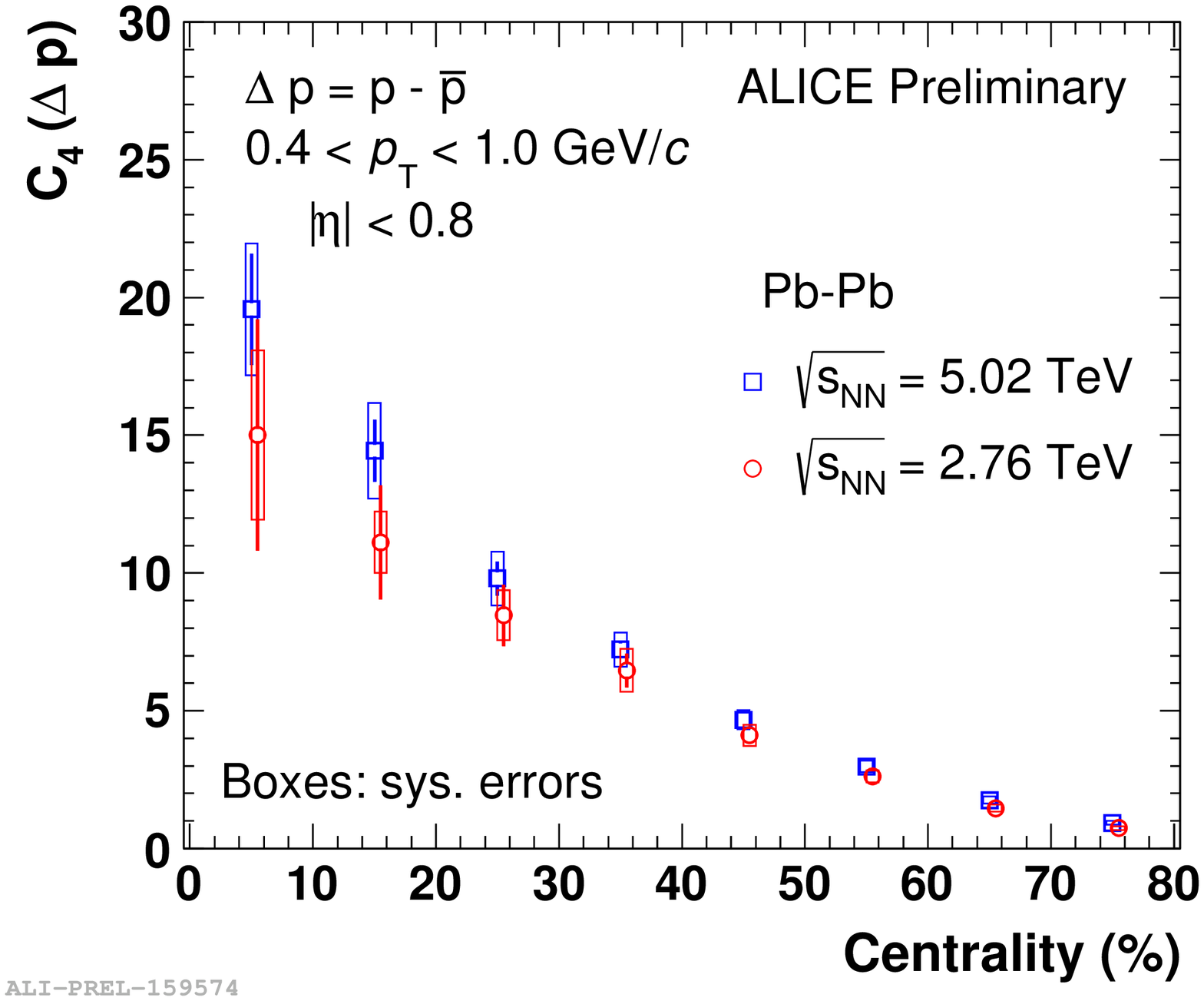}
\end{array}$
\end{center}
\caption{ (From left to right) Centrality dependence of $\mathrm{C}_{2}$, $\mathrm{C}_{3}$ and $\mathrm{C}_{4}$ of
  net-proton distributions in Pb--Pb collisions at
  $\sqrt{s_{\mathrm{NN}}}$ = 2.76 and 5.02 TeV. The the vertical lines
  and boxes represents the statistical and systematic uncertainties, respectively.}
\label{NetPCn}
\end{figure}

The second, third and fourth order cumulants of net-proton distributions in Pb--Pb collisions at
$\sqrt{s_{\mathrm{NN}}}$ = 2.76 and 5.02 TeV are shown in Figure
\ref{NetPCn} as a function of centrality. It can be observed from Figure \ref{NetPCn} that the $\mathrm{C}_{2}$ and $\mathrm{C}_{4}$ of net-proton distributions are decreasing from central to peripheral events, while $\mathrm{C}_{3}$ does not show a strong centrality dependence. These results are the same, within the uncertainties, for both the energies.

\begin{figure}[h]
\begin{center}$
\begin{array}{cc}
\includegraphics[width=2.0in]{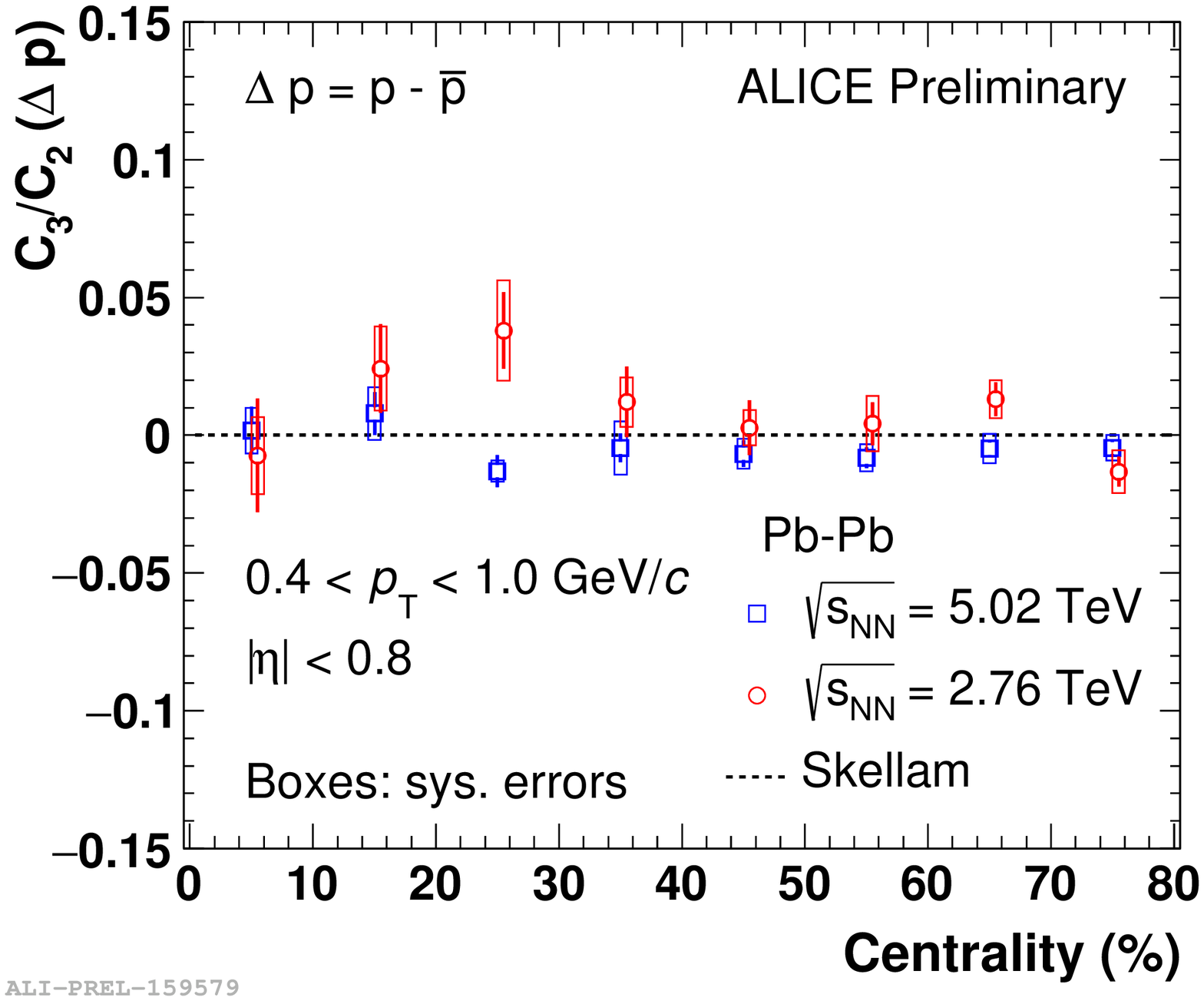} &
\includegraphics[width=2.0in]{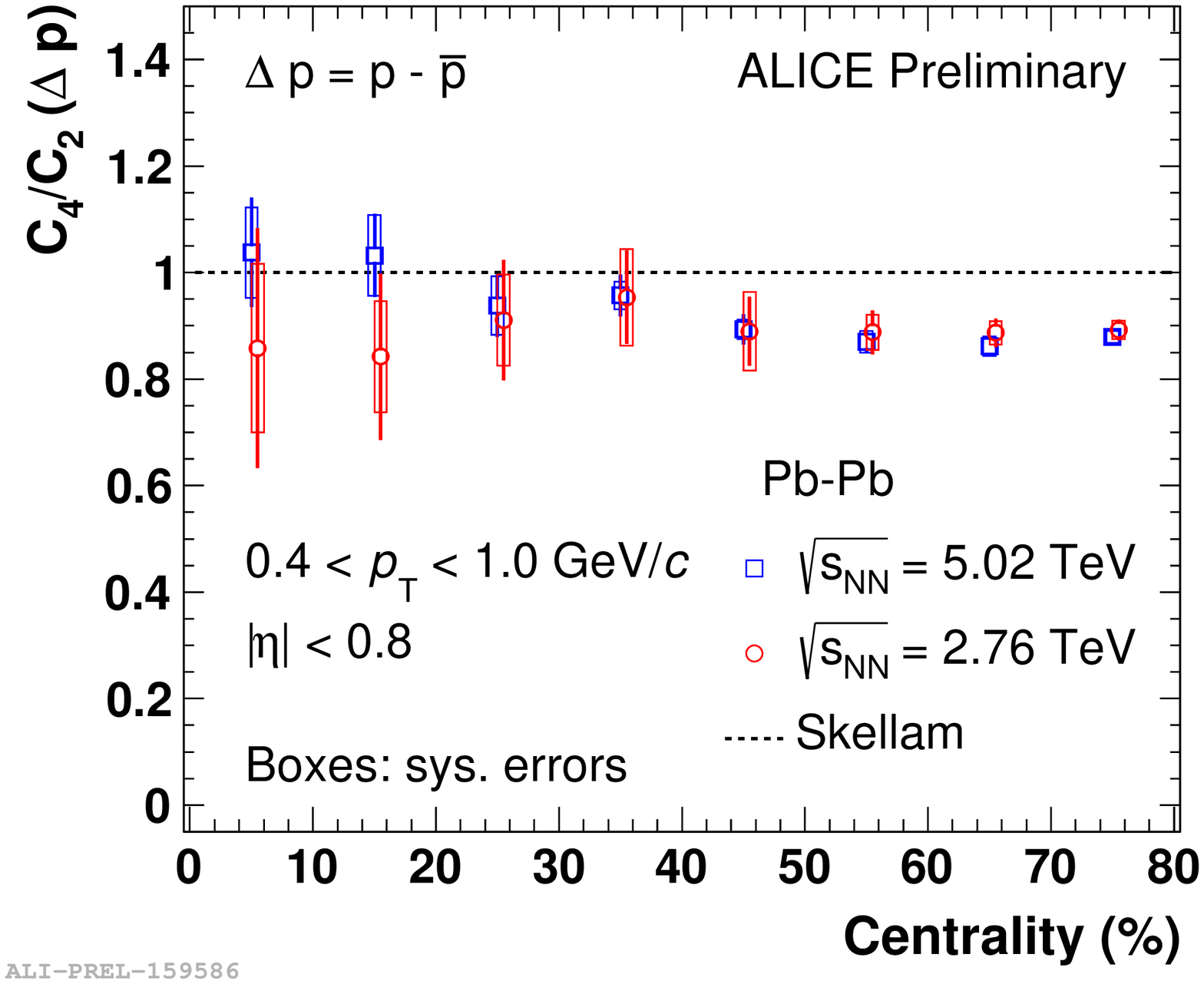}
\end{array}$
\end{center}
\caption{Centrality dependence of
  $\mathrm{C}_{3}/\mathrm{C}_{2}$ and $\mathrm{C}_{4}/\mathrm {C}_{2}$ of
  net-proton distributions in Pb--Pb collisions at
  $\sqrt{s_{\mathrm{NN}}}$ = 2.76 and 5.02 TeV. The the vertical lines
  and boxes represents the statistical and systematic uncertainties, respectively.}
\label{NetPCnRatio}
\end{figure}

In Figure \ref{NetPCnRatio}, $\mathrm{C}_{3}/\mathrm{C}_{2}$ and
$\mathrm{C}_{4}/\mathrm {C}_{2}$ of net-proton distributions in Pb--Pb
collisions at $\sqrt{s_{\mathrm{NN}}}$ = 2.76 and 5.02 TeV are shown
as a function of collision centrality. The results are compared with
Skellam expectations. If the numbers of protons and anti-protons are distributed according to two independent Poissonians, then the net-proton number will follow a Skellam distribution. The cumulants
$\mathrm{C}_{n}$ of a Skellam distributions are given as,
$\mathrm{C}_{n} = \mathrm{C}_{1}(\mathrm{p}) +
(-1)^{n}~\mathrm{C}_{1}(\mathrm{\bar{p}})$.  The dotted lines in
Figure \ref{NetPCnRatio} represent the Sekllam expectations. It can be
seen from Figure \ref{NetPCnRatio} that values of
$\mathrm{C}_{3}/\mathrm{C}_{2}$ are within the Sekllam expectations for energies in all centrality bins considering the uncertainties. In central and semi-central events, values of $\mathrm{C}_{4}/\mathrm {C}_{2}$ of
net-proton number fluctuations agree with Skellam expectations. However,
some significant deviations from the Skellam line is observed in
peripheral events, which will be the subject of future investigations.

\begin{figure}[h]
\begin{center}$
\begin{array}{cc}
\includegraphics[width=2.0in]{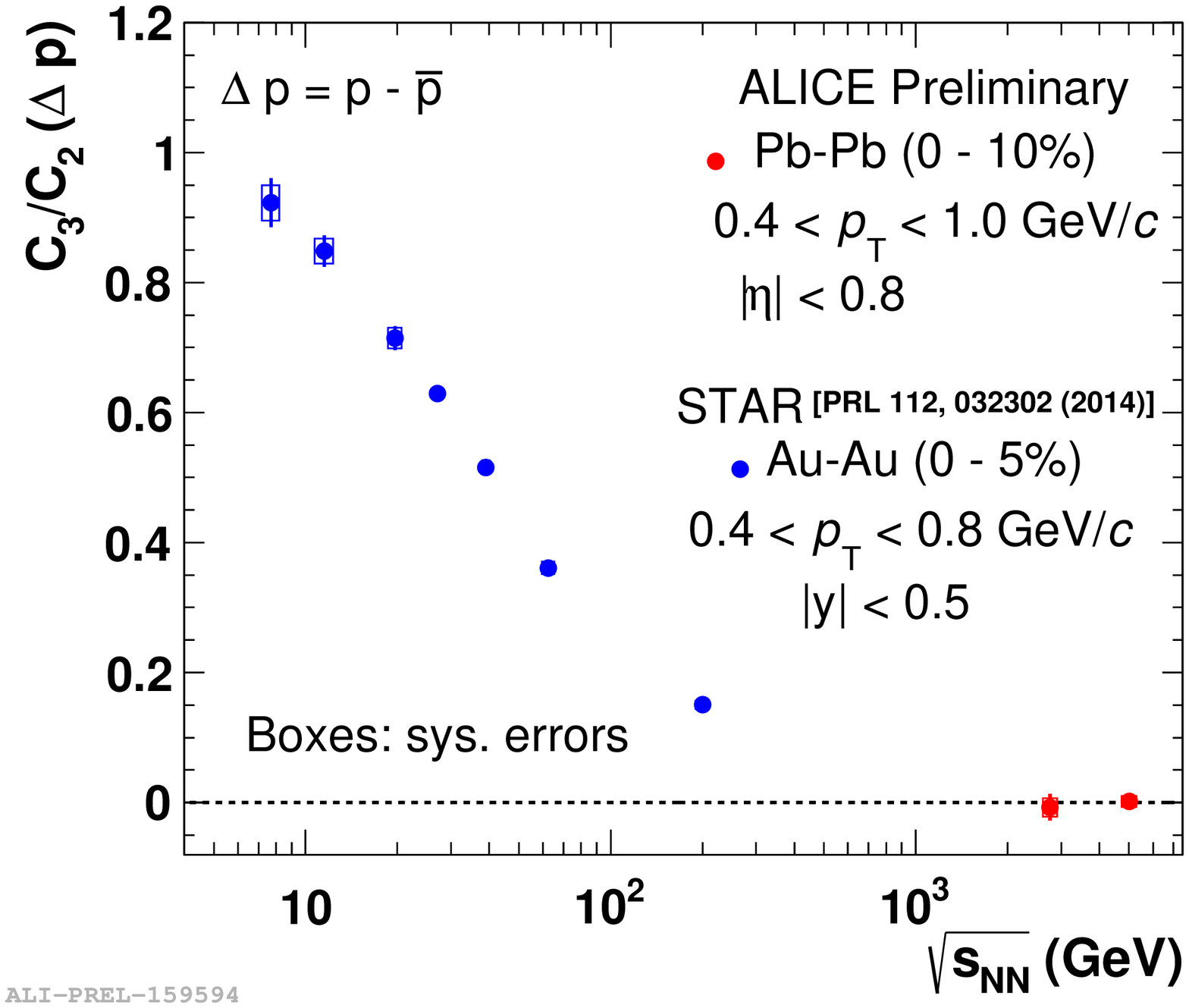} &
\includegraphics[width=2.0in]{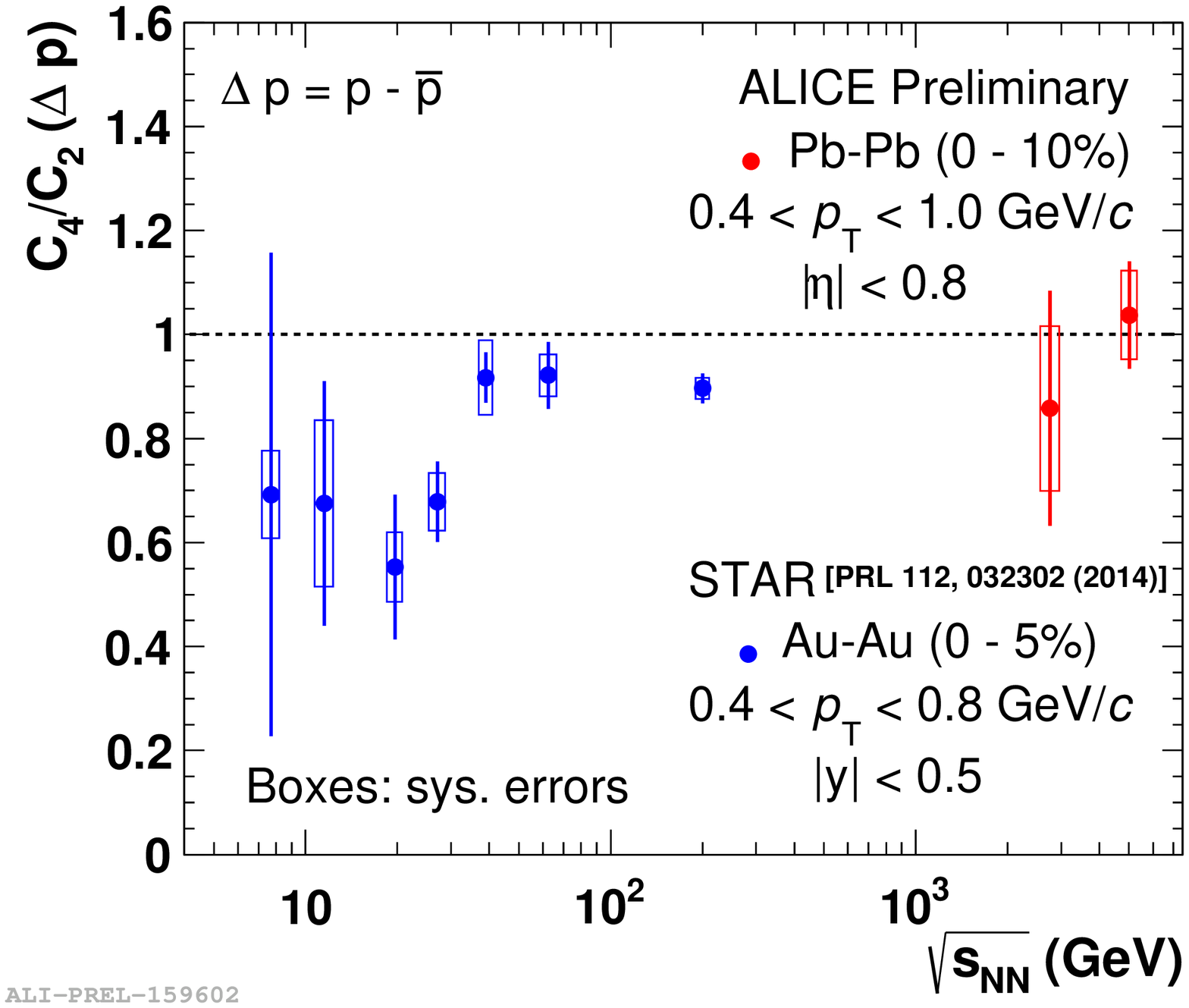}
\end{array}$
\end{center}
\caption{Energy dependence of
  $\mathrm{C}_{3}/\mathrm{C}_{2}$ and $\mathrm{C}_{4}/\mathrm {C}_{2}$ of
  net-proton distributions for the most central collisions.}
\label{NetPEnCn}
\end{figure}

We also compare our results with RHIC BES results of net-proton
measured in $0.4 < p_{T} < 0.8$ GeV/$c$ and $|y| < 0.5$ \cite{Adamczyk:2013dal}. The beam
energy dependent results of $\mathrm{C}_{3}/\mathrm{C}_{2}$ and
$\mathrm{C}_{4}/\mathrm {C}_{2}$ are shown for the most central
collisions in Figure \ref{NetPEnCn}. It should be noted that our
results are obtained in $0.4 < p_{T} < 1.0$ GeV/$c$ and pseudorapidity range 
$-0.8 < \eta < 0.8$. However, using the RHIC kinematic ranges has a minimal influence on our results. It can be seen that going
from RHIC to LHC energy, values of $\mathrm{C}_{3}/\mathrm{C}_{2}$ and
$\mathrm{C}_{4}/\mathrm {C}_{2}$ of net-proton number fluctuations approach Skellam expectations within the relatively small kinematic window. 

\section{Summary}
In these proceedings, we presented the first measurements of net-proton cumulants up to $4^{th}$ order and their ratios in Pb--Pb collisions
at $\sqrt{s_{\mathrm{NN}}}$ = 2.76 and 5.02 TeV as a function of centrality. Within
the current kinematic cuts, the cumulant ratios ($\mathrm{C}_{3}/\mathrm{C}_{2}$,
$\mathrm{C}_{4}/\mathrm {C}_{2}$ ) results are found
to be consistent with Skellam expectations within the
uncertainties. We also compared our results concerning the net-proton cumulants with those obtained at RHIC by the STAR experiment with the BES program. From RHIC to LHC, the ratio approaches the Skellam expectations. The upcoming dedicated Pb--Pb run at 5.02 TeV will improve the statistical precision to further constrain the freeze-out parameters. The study of higher order cumulants of net-charge and net-kaon in a wide kinematic region is also a subject of future investigations. 

\section{Acknowledgements}
This work was supported by National Research Foundation of Korea (NRF), Basic Science Research Program through the National Research Foundation of Korea funded by the Ministry of Education, Science and Technology (Grant number: NRF-2014R1A1A1008246).





\bibliographystyle{elsarticle-num}
\bibliography{<your-bib-database>}

\begin{thebibliography}{00}


\bibitem{Aoki:2006we} 
  Y.~Aoki, G.~Endrodi, Z.~Fodor, S.~D.~Katz and K.~K.~Szabo,
  Nature {\bf 443}, 675 (2006).


\bibitem{Friman:2011pf} 
  B.~Friman, F.~Karsch, K.~Redlich and V.~Skokov,
  Eur.\ Phys.\ J.\ C {\bf 71}, 1694 (2011).

\bibitem{Andronic:2009qf} 
  A.~Andronic, P.~Braun-Munzinger and J.~Stachel,
  Acta Phys.\ Polon.\ B {\bf 40}, 1005 (2009).

\bibitem{Andronic:2016nof} 
  A.~Andronic, P.~Braun-Munzinger, K.~Redlich and J.~Stachel,
  J.\ Phys.\ Conf.\ Ser.\  {\bf 779}, no. 1, 012012 (2017)

\bibitem{Bhattacharya:2014ara} 
  T.~Bhattacharya {\it et al.},
  Phys.\ Rev.\ Lett.\  {\bf 113}, no. 8, 082001 (2014).

\bibitem{Karsch:2010ck} 
  F.~Karsch and K.~Redlich,
  Phys.\ Lett.\ B {\bf 695}, 136 (2011).

\bibitem{Aamodt:2008zz} 
  K.~Aamodt {\it et al.} [ALICE Collaboration],
  JINST {\bf 3}, S08002 (2008).

\bibitem{Aamodt:2010cz} 
  K.~Aamodt {\it et al.} [ALICE Collaboration],
  Phys.\ Rev.\ Lett.\  {\bf 106}, 032301 (2011).

\bibitem{Wang:1991hta} 
  X.~N.~Wang and M.~Gyulassy,
  Phys.\ Rev.\ D {\bf 44}, 3501 (1991).
  
  \bibitem{Brun:1994aa}
  R.~Brun, F.~Carminati, and S.~Giani, CERN-W5013 (1994).


\bibitem{Nonaka:2017kko} 
  T.~Nonaka, M.~Kitazawa and S.~Esumi,
  Phys.\ Rev.\ C {\bf 95}, no. 6, 064912 (2017).
  
  \bibitem{Luo:2013bmi} 
  X.~Luo, J.~Xu, B.~Mohanty and N.~Xu,
  J.\ Phys.\ G {\bf 40}, 105104 (2013).
  
  \bibitem{Adamczyk:2013dal} 
  L.~Adamczyk {\it et al.} [STAR Collaboration],
  Phys.\ Rev.\ Lett.\  {\bf 112}, 032302 (2014).

\end{thebibliography}



\end{document}